\documentclass[twocolumn]{aastex7}

\begin{document}

\title{Fast Radio Bursts as Cosmic Lightning}

\author[orcid=0009-0003-9026-5802,sname='Kafashi']{Parsa Kafashi}
\affiliation{Department of Physics, Sharif University of Technology , Azadi ave. Tehran 11365-9161, Iran}
\email[show]{p.kafashi@gmail.com}  

\author[orcid=0000-0002-7084-5725,sname='Rahvar']{Sohrab Rahvar}
\affiliation{Department of Physics, Sharif University of Technology , Azadi ave. Tehran 11365-9161, Iran}
\email[show]{rahvar@sharif.edu}  



\begin{abstract}

We propose a new model for the origin of Fast Radio Bursts (FRBs), attributing these phenomena to sudden discharges of accumulated electric charge in the accretion disk of compact objects such as black holes. Our framework demonstrates how Compton scattering within the disk plasma generates charge separation, creating a capacitor-like system stabilized by the equilibrium between radiation pressure and electrostatic forces. We detail the discharge process through destabilizing mechanisms in this capacitor, resulting in radiative emission. We compare our model's prediction on radiation signatures with observational data, using FRB2018725A as an example to obtain key quantitative relationships. Additionally, we estimate the total charge buildup via Compton scattering for a stellar-mass black hole, constrained by the best-fit  between our model and observations, and determine the corresponding electron density in the accretion disk for this mechanism to operate.


\end{abstract}


\section{Introduction}\label{Introduction}

Fast radio bursts (FRBs) are intense, millisecond-duration radio signals originating from beyond our galaxy. They were first identified by \cite{2007Sci...318..777L}
 through archival data from the Parkes Observatory in Australia. These enigmatic transients have spurred numerous theoretical explanations \citep{popov2010, kulkarni2014, lyubarsky2014, Cai2018, Katz2016, Loeb2016, refId0}. Yet despite extensive study, their underlying physical mechanisms remain unresolved.


For instance, \cite{popov2010} proposes that FRBs could be produced by neutron stars in binary systems with an accretion disk, where magnetic activity or interactions with surrounding material generate the bursts. \cite{kulkarni2014} proposes that FRBs may originate in supernova remnants, specifically through interactions involving magnetars—neutron stars with extremely strong magnetic fields. In a different approach, \cite{lyubarsky2014} attributes the origin of FRBs to magnetic reconnection events in magnetars' magnetospheres, where magnetic field lines snap and release energy.

The dispersion measure (DM) of FRBs, which quantifies the observed frequency-dependent dispersive delays in terms of free-electron column densities, greatly exceeds predictions from models of the Milky Way interstellar medium, indicating an origin at cosmological distances. These bursts occur at a surprisingly high rate of thousands per day in all-sky surveys, making it a common cosmic phenomena associated with extreme environments and characterized by energy densities about ten billion times larger than those from galactic pulsars \citep{annurev:/content/journals/10.1146/annurev-astro-091918-104501}. While non-repeating FRBs constitute the majority of observed events, a subset of FRBs exhibit the remarkable characteristic of repetition. The repetitive bursts have different repetition times and complex patterns, making it unclear whether some are a one-time-only phenomenon or if all FRBs are repetitive, but with vastly different repetition times. Despite the rarity of repeating FRBs in comparison, these FRBs have garnered significant attention from the scientific community due to their potential to provide insights into the underlying mechanisms responsible for these enigmatic cosmic phenomenon. It is not clear whether the repeating and non-repeating FRBs have the same nature \citep{10.1093/mnras/staa1361}.

In this study, we propose a novel mechanism for fast radio burst  generation through a discharge process in a naturally forming capacitor-like system within accretion disks around compact objects. Our model demonstrates how radiation pressure in the disk plasma displaces free electrons, inducing charge separation and effectively "charging" this astrophysical capacitor. When disk instabilities disrupt the equilibrium state, the system undergoes rapid discharge, releasing stored energy as coherent radio emission consistent with observed FRB properties.



 In section \ref{Mechanism}, we provide a comprehensive explanation of the functioning of our proposed mechanism for FRBs, including the radiation process in this model. In Section \ref{Radiation from the discharge} we calculate the radiation process during the discharge of natural capacitor. 
 In Section \ref{sec:Comparison with observed data}, we compare our model with observational data and analyze FRB2018725A. By fitting the radiation pattern from our model to the light curve of the observed bursts, we derive numerical values for the model parameters.

 In Section \ref{Calculating the charge}, we derive the characteristics of the capacitor-like system and the conditions of its surrounding environment.

 Finally, in Section \ref{sec:conclusions}, we summarize our work and discuss possible future directions.

\section{Mechanism} \label{Mechanism}

In the accretion disks around compact objects, the plasma may be considered collisionless. However, the interaction of photons with this medium can have different effects on the electrons and ions. Moreover, the cross section for photon-electron interaction is higher than the ions. With high energy photons in the accretion disk, their interaction with the plasma results in the spatial separation of electrons and ions due to their considerable mass difference as well as the difference in their scattering cross section.
In this process, electrons are propelled outward by Compton scattering, causing them to separate from the heavier ions. The conditions that must be met for this separation to occur are studied in \citep{Del_Gaudio_2020}.

Here we consider a radiation source at the center of an accretion disk emitting high energy photons. As a toy model, this leads to charge separation on the disk, with a negatively charged layer consisting of the scattered electrons on the outer layer, and a positively charged layer consisting of protons and ions in the inner layer. The charge accumulation in plasma through Compton scattering is shown in experiments.  \cite{Gong_2017} shows a similar process, in which the electrons in the plasma are boosted through direct laser acceleration while the ions are driven inward as a result of laser ponderomotive force. The accumulation of charge will continue until an equilibrium is reached in the forces acting on the charged particles. As the charge in the layers increases, so does the electric force between the inner and outer layers. Once this force is great enough to counteract the other forces acting on the particles, such as gravity and radiation pressure, there will no longer be any accumulation of charge in the layers and the system will reach a stable configuration.

Due to the chaotic nature of the accretion disks, the system will experience radiation fluctuations \citep{rapidspinchange}, leading to the collapse of the otherwise stable configuration. When a disturbance in the radiation pressure occurs, the electrons in the outer layer can fall due to the attraction to the inner layer protons. This extreme movement similar to that of a lightning discharge but on vastly larger scales, can cause an immense burst of radiation, which we model to be the source of the fast radio bursts.

To study this phenomenon we first model a stable configuration for the capacitor, and the sudden collapse to the discharge of this capacitor. Details on the mechanism for the charge separation will be discussed later. 
\section{Radiation of the discharge} 

\label{Radiation from the discharge}
As previously mentioned, the charge formation can be modeled as a capacitor and the discharge can be taken as the capacitor's discharge in an RC circuit  (Resistance-Capacitor), with the resistance being the medium in between the layers, consisting of plasma.

During the discharge of the capacitor, currents are more likely to flow through specific regions rather than uniformly across the entire layers. In other words, the current density between the two layers is not homogeneous, leading to localized discharges at particular points
By this assumption, we can take the current as $I(t)={Q}\omega e^{-\omega t}$ for the discharge, where $1/\omega$ defines the time-scale for the RC circuit as $T = 1/\omega = RC$. From this electric current, the electromagnetic potential for the two charged layers with distance $l$ from each other, at the distance of $r$ from the source is
 \begin{equation}\label{s}
            \vec{A}(r,t) =\frac{\mu_0}{4\pi}\frac{Q\vec{l}\omega e^{-\omega t}}{r}.
 \end{equation}

 The radial component of the Poynting vector for this source at far field approximation is
 \begin{equation}\label{eq2}
S_r(r,\theta,\phi,t)=\frac{Q^2l^2}{32\pi^2T^4\epsilon_0c^3} \frac{\sin^2{\theta}}{r^2}e^{-\frac{2t}{T}}, 
 \end{equation}
 where we have taken the space in between the source and the observer as vacuum and the $\epsilon_0$ is the electric permeability of the vacuum. 

 Taking into account the solution of Maxwell equation in FLRW metric,  the Poynting is modified compared to the Minkowski space. 
 To do so, we  define the conformal time $\eta$, to transform the FLRW metric to a conformally flat metric with conformal scale factor of $a(\eta)$. Then for a flat universe (i.e. flat in the spatial component of metric) we find the radial component of the Poynting vector received by an observer at comoving distance of $r$ as
 \begin{equation}   \label{theoretical flux}
    S_0(t)= \frac{Q^2l^2}{32\pi^2T^4\epsilon_0c^3} \frac{\sin^2{\theta}}{d_L^2}e^{-\frac{2t}{T}\left(1+\left(\frac{r}{3tc}\right)\right)^3},
\end{equation}
where $d_L=a(t_0)r(1+z)$ is the luminosity distance, $z$ is the cosmological redshift and we will set $a(t_0)=1$. (For details see Appendix \ref{appendix A}).\\

\section{Comparison with observed data} \label{sec:Comparison with observed data}
From the data provided by \cite{2021}, we use our theoretical prediction in equation (\ref{theoretical flux}) to compare with the observed data. From the observed data we estimate relevant physical parameters for FRBs.
Since the data provides the flux density per unit frequency in units of Jansky,  the total flux in terms of the flux density is
\begin{equation}
    S(t)=\int_{\nu} S_\nu d\nu,
\end{equation}
where we can take integration within the range of 
 observing band of $\nu_{lo}$ and $\nu_{hi}$. For simplicity we adapt a reasonable estimate of the flux density $S_\nu$, by assuming it to be constant over the band.  Under this approximation we have $S(t)\simeq S_\nu$ within the range of $\nu_{lo}<\nu<\nu_{hi}$.
 
The signal coming from a distant source experiences dispersion and scattering along its way, characteristic of the medium through which it travels. The scattering causes each initial frequency component to spread out, approximated by a normal distribution around the initial frequency, which results in $S_\nu$ being wider and more distributed along frequencies. However, by integrating through frequencies and a sensible choice of $\nu_{lo}$ and $\nu_{hi}$, the scattering effect can be negligible on the appropriate $S(t)$. The signal can also experience broadening in time, due to each frequency traveling through the medium at different speeds and experiencing dispersion, which is already accounted for in the data.

The distance of the source of the FRB and its intrinsic luminosity can also be derived from observed values through the dispersion measure. The dispersion measure is given by
\begin{equation}
    DM=\int^{d}_{0}n_e(l)dl,
\end{equation}
where $n_e(l$) is the electron number density on the path length $l$  and $d$ is the distance to the FRB. The dispersion quantifies the time delay of the pulse between highest and lowest radio frequencies of observation. Therefore, we can find the value of the dispersion measure by the time delay of different frequencies in the observed data. For  FRB20180725A the dispersion measure is $715.8 \mathrm{pc\cdot cm^{-3}}$. Having the dispersion measure and the commonly used estimate of redshift to be $z < \frac{DM}{1000\mathrm{cm}^{-3}\mathrm{pc}}$, one can calculate the luminosity distance and source luminosity. \\
From \cite{Petroff_2019} the luminosity distance is
\begin{equation}
  d_L<\left(\frac{DM}{500\mathrm{cm}^{-3}\mathrm{pc}}\right)\left[\left(\frac{DM}{1000\mathrm{cm}^{-3}\mathrm{pc}}\right)+2.4\right] \mathrm{Gpc},   
\end{equation}
and considering an  isotropic equivalent source \citep{hogg2000distancemeasurescosmology}, the luminosity is 
\begin{equation}
    L_0=\frac{4\pi d_L^2S_{\nu}\Delta{\nu}}{\left(1+z\right)}
\label{L}
\end{equation}

\begin{figure}[h]
    \centering
    \includegraphics[width=1\linewidth]{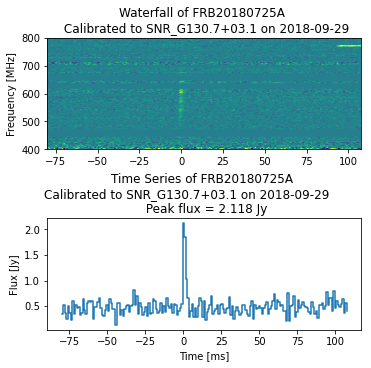}
    \caption{Calibrated time series of FRB2018725A and the graph representation of the flux density in time. The peak flux for this burst is 2.118 Jy with a fluence of 4.1 Jy.ms and  its frequency band between 485.3 and 760.1 MHz. This figure was adapted from \cite{2021}
    \label{fig:enter-label}}
\end{figure}

\begin{figure}
    \centering
    \includegraphics[width=1\linewidth]{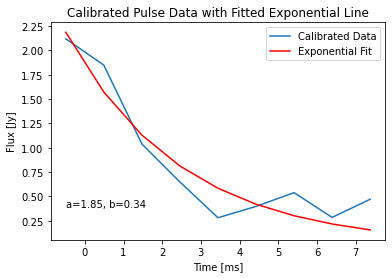}
    \caption{The original pulse data with its flux in units of Jansky fitted with an exponential curve as $ae^{-bt}$. The evaluate $T$ from the best fit is around $5.88\times 10^{-3}$s. The spectrum has a peak in the frequency of 607.4 MHz.}
    \label{fig:enter-label2}
\end{figure}

Figure (\ref{fig:enter-label}) shows the calibrated data for the FRB2018725A \citep{2021}. From the dispersion measure for this event, the upper limit for the redshift is $z<0.7$ which corresponds to $d_L<4.46 \mathrm{Gpc}$. In equation (\ref{L}), using the overall flux we obtain the intrinsic intensity of the FRB source as $L_0<8\times10^{35} \mathrm{W}$ and the total energy release as $E<7\times 10^{33}J$. We note that we took into account an isotropic radiation, so the real value should be smaller than this.
In the next step, we take the profile of flux as function of time (i.e. $S(t)$) and compare it with our model. 
Figure (\ref{fig:enter-label2}) represents the best fit of an exponential curve $a e^{-bt}$ to the calibrated FRB data \citep{Andersen_2023}.
From the best first the time-scale for this burst is  $T=1/\omega = 5.88\times10^{-3}\mathrm{s}$. One the other hand from the value of $a$ parameter we extract extra information:

\begin{equation} \label{p2}
    \frac{Q^2l^2\sin^2{\theta}}{32\pi^2\epsilon_0c^3T^4d_L^2}=a
\end{equation}
Using equation(\ref{p2}) and the numerical value of $T$ and the distance, averaging over all angles we estimate the dipole term in this model,
\begin{equation}
\label{psin}
    Ql\simeq 1.3\times 10^{30} \mathrm{C\cdot m}
\end{equation}

\section{Calculating the charge} \label{Calculating the charge}

Our next step is to calculate the charge accumulation in this RC model.
In the accretion disk, the high energy photons collide with the free electrons in the plasma causing a Compton scattering. The propelled electrons will then be scattered in the accretion disk, creating a negatively charged layer.
We start with calculating the rate of interaction between the incoming photons and the free electrons in the plasma for a volume with  an area of $A$ and width of $x$. Considering the scattering starts from a distance of $s$, the number of interactions per time would be
\begin{equation}\label{number of interaction per time}
    \frac{dN_I}{dt}=\int_s^{s+x}\frac{LA}{4\pi r^2h\nu}n_e \sigma_e dr
\end{equation}
where $L$ is the luminosity, $\sigma_e$ is the electron cross section (see Appendix 2) , $\nu$ is the frequency of the photons and $n_e$ is the electron density. We have taken the electron density to be a constant. One can extend to a power-law  density profile \citep{Abdikamalov2021_RELXILLDGRAD_NK}, However, we have checked that this modification makes negligible difference in our results . Moreover, in the integral of equation (\ref{number of interaction per time}), a factor of $e^{-\tau (r)}$ where $\tau$ is the optical depth defined as:
\begin{equation}
    \tau(r)=\int_s^{s+r} n_e(r^{\prime}) \sigma_e dr^{\prime}, 
\end{equation}
should be included to account for the absorption of photons as they propagate through the medium. Nevertheless, we neglect this factor here, 
 as we will later take $x$ in a way that $\tau $ doesn't exceed one.

Having the interaction rate, we can calculate the radiation pressure by:
\begin{equation}
\label{rpressure}
    P_r= \frac{dN_I}{Adt}\Delta p,
\end{equation}
where $\Delta p$ is the change in momentum of an electron due to interaction with photons which can be taken as the maximum momentum change of an electron in Compton scattering.
\begin{equation}
    \Delta p=m_e c \frac{2\alpha(1+\alpha)}{1+2\alpha},
\end{equation}
where $\alpha = \frac{h\nu}{m_ec^2}$.

After the scattering of electrons, the electrons move in the plasma and make a negatively charged layer of electrons at the outer layer while the inside layer has positive charge made of protons and ions.  These two layers of electric charge can act as plates of a capacitor, which experience electric pressure. Having the electric pressure and the radiation pressure exerted on the layers in equilibrium, result in a state with an accumulated charge $Q$ on the plates. 


The electric pressure on the plates is
$P_e =  \sigma^2/2\epsilon$, where $\sigma$ is the surface charge density of the layers.
Equating with equation (\ref{rpressure}), results in 
\begin{equation}
  \frac{\sigma^2}{2\epsilon}=\frac{L\sigma_cn_e}{4\pi c}\left(\frac{1}{s}-\frac{1}{s+x}\right)\left( 1+ \frac{1}{1+2\alpha}  \right).
\end{equation}
Then the critical charge accumulated on the outer layer would be
\begin{equation}\label{bigQ}
    Q=4\pi s^{\prime 2}\left(\frac{L\sigma_c\epsilon n_e}{2\pi c}(\frac{1}{s}-\frac{1}{s+x})( 1+ \frac{1}{1+2\alpha}  )\right)^{1/2}
\end{equation}
Where $s^{\prime}$ is the negatively charged layer's radius from the center. This is an expression for the accumulated charge.

In order to estimate a value for $Q$, we take a $10$ stellar-mass blackhole at the center of accretion disk.
Then we use the Eddington luminosity to estimate the luminosity of photons in our model where for a blackhole with $M = 10 M_\odot$, the luminosity is $L = 10^{32}$ W \citep{eddington2002apa..book.....F}, noting that this luminosity is different from of FRB luminosity, $L_0$, in our model. One the other hand, we assume that the scattering of photons start from the innermost stable circular orbit (ISCO), which is a few times bigger than the Schwarzschild radius of the blackhole \citep{Zajaček_2019}. The ISCO of a typical stellar-mass blackhole would be around $10^5 \mathrm{m}$. We use this value in the integration of equation (\ref{number of interaction per time}).

 We argue that the characteristic width $x$ is of the order of the mean free path of photons in the plasma. This is justified by considering that the typical energy of photons produced via Bremsstrahlung is significantly higher than the thermal energy of electrons in the plasma. Consequently, after undergoing a single interaction over a photon mean free path, the photon energy is not sufficient to displace electrons further outward. While the electron mean free path due to Coulomb scattering is much shorter than that of photons (Appendix \ref{appendix 2}), the net outward flow of photons from the inner regions of the accretion disk effectively acts like an electric field, exerting a drift force on the electrons.

 In order to make a circuit-model for the motion of electrons inside the plasma, let us recall that the mean free path of electrons is short. Similar to the electric field in the circuit that pushes the electrons inside the metal, here we replace the battery with the momentum transfer of the photon radiation from the inner parts of the accretion disk. The radiation of disk plays the electromotive force and similar to analyzing a circuit, we have to check if the momentum transfer to the electrons is enough to have constant drift velocity before thermal dissipation due to interaction with the other electrons.

To satisfy this condition we compare the two time scales of the electron scattering with another charge inside the plasma $t_e$, with the time-scale of interaction of the photons radiating from the disk with the electrons inside the plasma $t_\gamma$. Having the condition of $ t_\gamma  \ll t_e$, guarantees that consecutive photon momentum transfer will push the electrons and move them globally inside the plasma. However, after one mean free path of the photons inside the plasma, the photons lose their momentum and no more momentum transfer happens to the electrons.

In what follows we compare these two time scales.  The collision time which is based on Coulomb scattering for a relativistic electrons is (Appendix~\ref{appendix 2}):
\begin{equation}
    t_e = \frac{(4\pi\epsilon_0)^2 m_e^2 v^3 \gamma^2}{8\pi n_e e^4 \ln{\Lambda}},
    \label{collision frequency}
\end{equation}
where \( v \) is the velocity of electrons, \( \gamma \) is the Lorentz factor, \( n_e \) is the electron number density, and \( \Lambda \) is the plasma parameter.

On the other hand, we can obtain the time scale that a single electron is scattered by photons emitted from the inner parts of the accretion disk. 
At a radial distance \( r \) from the source, within a differential depth \( dr \), the number of electrons is \( dN_e = n_e (2\pi H r\, dr) \), where \( H \) is the thickness of the accretion disk. Substituting in equation (\ref{number of interaction per time}), the rate of photon-electron interactions is:
\begin{equation}
    \frac{d^2N_I(r)}{dtdr} = \frac{2\pi r H L\, n_e\, \sigma_e\, e^{-\tau(r)}}{4\pi r^2 h\nu},
\end{equation}
where \( L \) is the luminosity, the flux is distributed over an area \( 4\pi r^2 \), and the target electron area is \( 2\pi r H \). Here, \( \sigma_e \) is the scattering cross-section, \( \nu \) is the photon frequency, and \( \tau(r) \) is the optical depth.

Therefore, the average time it takes for a single electron to be struck by a photon is:
\begin{equation}
    t_{\gamma} = \frac{\frac{dN_e}{dr}}{\frac{d^2N_I}{dtdr}} = \frac{4\pi r^2  h\nu}{L \sigma_e e^{-\tau(r)}}.
\end{equation}

And consequently, the ratio of the photon-electron interaction timescale to the electron-electron collision timescale is:
\begin{equation}
    \frac{t_{\gamma}}{t_e} = 
    \frac{2r ^2\, h\nu\, n_e\, e^4 \ln{\Lambda}}{L\, \sigma_e\, e^{-\tau(r)}\, \epsilon_0^2\, m_e^2\, v^3\, \gamma^2}.
\end{equation}

\begin{figure}
    \centering
    \includegraphics[width=1\linewidth]{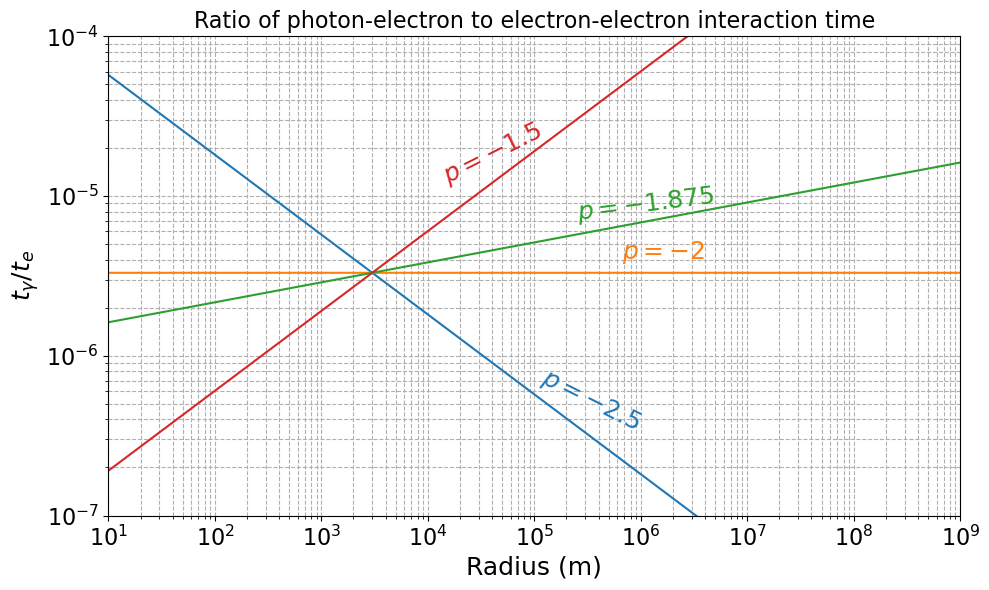}
    \caption{The ratio of the average time it takes for an electron to be scattered by a photon, to the collision time of a relativistic electron, for an electron density profile \( n_e = n_0 \left(\frac{r}{r_0}\right)^p \). Where $n_0$ is $10^{18}\mathrm{m}^3$ and $r_0$ is $3\times10^3 \mathrm{m}$}
    \label{fig:ratio of times}
\end{figure}

This ratio is plotted in Figure~\ref{fig:ratio of times} for different power-law indices of the electron density profile. Since \( t_{\gamma} \ll t_e \), the timescale for photon interactions is much shorter than that of electron-electron collisions. Thus, electrons are more frequently impacted by photons and tend to be swept outward over a distance comparable to the photon mean free path.

 In order to calculate the accumulation of charge at mean free path distance of photons, we use the numerical values for the number density of electrons and Compton cross section (see Appendix B), the optical depth is around $x \simeq 10^{10}$m.

Substituting in the equation \ref{bigQ},  we estimate the theoretical critical charge $Q\simeq 10^{20}$C. We note that this is a preliminary estimation for the charge of model. In order to fix the parameter of the accretion disk, we need to compare the observational value derived from the luminosity with the theoretical model. 

\begin{figure}[t]
    \centering
    \includegraphics[width=1\linewidth]{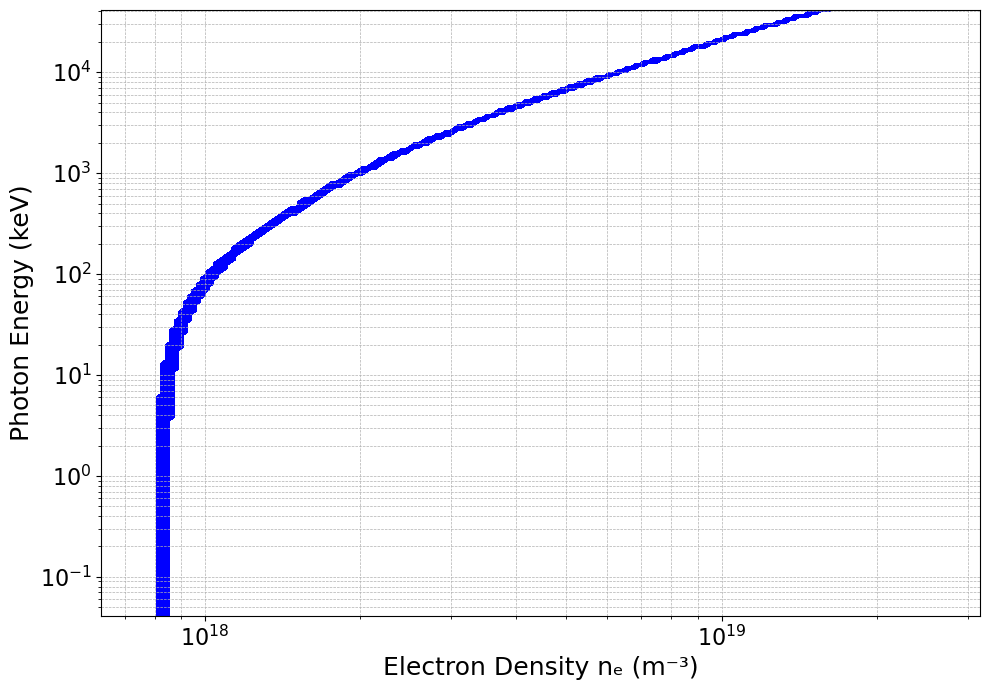}
    \caption{The matching condition between the theoretical charge and the charged derived from observational data . The plot shows the relationship between the electron density in the accretion disk and the photon energy required for the model to be consistent with the observational data. The possible values are calculated in a way that the theoretical charge would within one percent of the charged derived from observational data.}
    \label{loglog Qq}
\end{figure}

Let us take $Q$ as the charge of model that we can derive from the observation. From the observation we have $T $ as the damping time-scale of FRB as well as the dipole from equation (\ref{eq2}). Moreover if we identify the accretion disk that sources FRB, then we can measure the X-ray emission from the disk. Now we combine equation (\ref{bigQ}) with equation (\ref{psin}) to constrain the number density of electrons and the frequency of X-ray radiation from accretion disk. 
Figure (\ref{loglog Qq}) shows this constrain. This range from the electron density and the photons energy is reasonable for the accretion disk X-ray sources. For instance taking $10\mathrm{keV}$ (\cite{García_2010}), then the corresponding electron density of the plasma would be on the order of $10^{18}\mathrm{m}^{-3}$ and the resulting critical charge in the RC model would be approximately $10^{20}\mathrm{C}$.

\section{Conclusions} \label{sec:conclusions}

We provided a model for FRBs generation in an X-ray accretion disk around a compact object. The X-ray photons generated in the accretion disk can 
scatter electrons, pushing them up to the distance of the mean free path of the photons. The result is a layer of accumulated charge which acts as a capacitor with its counterpart layer of positive charged ions. The continuous collision of the radiation makes this separation of changes much larger than the mean free path of electrons, which is similar to imposing a battery to an RC model.

While the radiation pressure and the electrostatic pressure make this capacitor stable, any perturbation in this system causes discharging of the capacitor and results in radiation in radio wave band. We obtained the profile of this radiation, exponentially decreasing with the characteristic damping time equal to the capacitance times the ohmic resistance of this model. We compared our model with the observational data. Taking the light curve of FRB2018725A we calculated the damping time scale as well as the dipole moment of this system as $10^{30} \mathrm{C\cdot m}$. On the other hand from the optical depth of the photons which is about the distance between the plates of the capacitor, we derived the charge on the capacitor in the order of $q \simeq 10^{20}$C. Moreover the model provides us a range for the density of electrons and the energy of X-ray photons from the accretion disk (See Figure \ref{loglog Qq}).



Comparing the observational data with this toy model for the FRBs provides reasonable values for the parameters of the model as the density of electron in the accretion disk and X-ray radiation from the disk. Further complication of the model is feasible as taking into account various disk models as well as the  spin of the disk, the magnetic field effect and turbulence in the disk that may explain some other feature of FRBs as repeating events.





\appendix

\section{Poynting vector in FLRW metric }
\label{appendix A}
The Maxwell equation in curved spacetime is
\begin{equation}
    \nabla_\nu F^{\mu\nu} = \mu_0 J^\mu,
\end{equation}
where in terms of the electromagnetic potential, 
\begin{equation}
    \nabla_\beta \nabla^\alpha A^\beta - \nabla_\beta \nabla^\beta A^\alpha = \mu_0 J^\alpha.
\end{equation}
Using the definition of Riemann tensor in terms of electromagnetic potential,  
\begin{equation}
    \nabla_\mu \nabla_\nu A_\rho - \nabla_\nu \nabla_\mu A_\rho = R^\alpha_{\ \rho \nu \mu} A_\alpha,
\end{equation}
where $ R^\alpha_{\ \rho \nu \mu} $ is the Riemann curvature tensor. We can rewrite Maxwell's equation as:
\begin{equation}
    \nabla^{\alpha} \nabla_{\beta} A^{\beta} + R^{\alpha}_{\ \rho} A^{\rho} - \nabla_{\beta}\nabla^{\beta} A^{\alpha} = \mu_0 J^{\alpha}, 
\end{equation}
where $R^{\alpha}_{\ \rho}$ is the Ricci tensor. 
We set the Lorenz gauge of  $\nabla_{\mu}A^{\mu}=0$, reduces the Maxwell equation to:
\begin{equation}
    R^{\alpha}_{\rho} A^{\rho} - \nabla_{\beta}\nabla^{\beta} A^{\alpha} = \mu_0 J^{\alpha}. 
\end{equation}

Now, we take the FLRW metric as the background metric, 
\begin{equation}
    ds^2 = -dt^2 + a^2(t) \left(dx^2 + dy^2 + dz^2 \right).
\end{equation}
We assumed spatially flat metric supported by the observation. For simplicity we introduce the conformal time by the definition of $d\eta={dt}/{da}$, the metric can be written as conformally flat metric 
\begin{equation} \label{Conformal metric}
    ds^2 = a^2(\eta)\left(-d\eta^2 + dx^2 + dy^2 + dz^2\right).
\end{equation}

The solutions of Maxwell equation in this metric compared to the Minkowski space is 
\begin{equation}
    \tilde{F}_{ab} = F_{ab}, \quad \tilde{F}^{ab} = a^{-4} F^{ab}, \quad \tilde{F}^b_a = a^{-2} F^b_a 
\end{equation}
\begin{equation}
    \tilde{A}_b = A_b, \quad \tilde{A}^b = a^{-2} A^b
\end{equation}

and the conformally rescaled Maxwell equations read:
\[
\tilde{g}^{ac} \tilde{\nabla}_c \tilde{F}_{ab} = -4\pi \tilde{j}_b
\]
\[
\tilde{\nabla}_{[a} \tilde{F}_{bc]} = 0, 
\]
provided that the four-current transforms according to:
\[
\tilde{j}_b = a^{-2} j_b, 
\]
Where the ones with tilde and the ones with no tilde represent the fields in the conformal metric and the Minkowski metric respectively. A detailed derivation is given in \cite{C_t__2019}. \\
The transformation of the stress-energy tensor is as:
\begin{equation}
    \tilde{T}_{ab} = a^{-2} T_{ab}, \quad \tilde{T}^{ab} = a^{-6} T^{ab}, \quad \tilde{T}^b_a = a^{-4} T^b_a
\end{equation}
So the Poynting vector which is $T^{0i}$ scales as:
\begin{equation}\label{conformal scaling of P vector}
    \tilde{S}^i(\eta)=a^{-6}S^i(t).
\end{equation}
Now that we have found the transformations of different expression in the conformally flat metric, we can find the relation of these expressions, in the source and observer frame. The propagation of a photon moving towards us in the conformally flat metric is governed by the following relation:
\begin{equation}
    \int_{\eta}^{\eta_0}d\eta=-\int_r^0 {dr}.
\end{equation}
so,
\begin{equation}\label{conformal time relation}
    \eta_0=\eta+{r}.
\end{equation}
So the coordinate time of $t$ will relate to the comoving time in the following way
\begin{equation}
\int^{\eta+\frac{r}{c}}_{\eta}d\eta=\int_{t}^{t_0}\frac{dt^{\prime}}{a(t^{\prime})},
\end{equation}
considering $a(t)=\left(\frac{t}{t_0}\right)^{\frac{2}{3}}$, we would have
\begin{equation}
    t=t_0\left(1-\left(\frac{r}{3t_0c}\right)\right)^3,
\end{equation}
So the Poynting vector for an observer at the comoving distance of $r$ that experiences both a dilation in the rate of arrival of photons and a decrease in the energy of the photons, each by a factor of $\frac{a(t)}{a(t_0)}=\frac{1}{1+z}$, is
\begin{equation}
        S_0(t) = \frac{Q^2l^2}{32\pi^2T^4\epsilon_0c^3} \frac{\sin^2{\theta}}{d_L^2}e^{-\frac{2t}{T}\left(1+\left(\frac{r}{3tc}\right)\right)^3}
\end{equation}
Where $d_L=a(t_0)r(1+z)$ is the luminosity distance.
In which the factor $a(t_0)$ comes from the fact that we have reverted back the space coordinates from the Minkowski metric to the FLRW metric for the observer $r\to a(t_0)r$. 
\\

\section{Mean free path of photons and electrons}\label{appendix 2}
From \cite{KleinNishina1929}, we have the general formula for differential cross-section:
\begin{equation}
\label{differential ccross section}
    \frac{d\sigma_e}{d\Omega} = \frac{3}{16\pi} \sigma_T \left( \frac{\epsilon_f}{\epsilon_i} \right)^2 \left( 1 + \frac{\epsilon_f}{\epsilon_i} - \sin^2 \theta \right)
\end{equation}
where $\sigma_T$ is the Thomson cross-section, \( \epsilon = h\nu \) is the photon energy, and \( \epsilon_f = \frac{\epsilon_i}{1 + \frac{\epsilon_i}{mc^2} \left[ 1 - \cos \theta \right]} \) is the final photon energy after scattering.\\
Integrating \ref{differential ccross section} over solid angle we will have the total Compton cross-section
\begin{equation}
    \sigma_{e} = \frac{3}{4} \sigma_T \Bigg[ \left( \frac{1 + \alpha}{\alpha^2} \left( 
\frac{2(1+\alpha)}{1+2\alpha} 
- \frac{\ln(1+2\alpha)}{\alpha} 
\right) \right) \nonumber  + \frac{\ln(1+2\alpha)}{2\alpha} 
- \frac{1+3\alpha}{(1+2\alpha)^2} 
\Bigg]
\end{equation}

Where $\alpha$ is $\frac{h\nu}{m_ec^2}$. \\
The optical depth $\tau_e$, which measures the probability for a photon to experience scattering in a region with electron density $n_e$ along the path length $x$ is defined as:
\begin{equation}
    \tau_e=\int_0^xn_e\sigma_edx
\end{equation}
From Beer-Lambert law we can find the ratio of the photons that experience scattering to all the photons to be $1-e^{-\tau_e}$ given as a function of $\tau_e$. The expected optical depth is then computed by $\int_0^\infty\tau_ee^{-\tau_e}d\tau_e$ which equals 1. This means that on average, a photon will experience scattering at a length of $\frac{1}{n_e\sigma_e}$ and this value will be taken as the width $x$ in the main body of the article.\\

To calculate the mean free path of an electron with relativistic speeds, through a plasma with electron density $n_e$, we will first find the average time it takes for a scattered electron to experience collisions. From \cite{nicholson1983plasma}, we know that the deflection of a charged particle in a plasma is predominantly  due to many random small angle collisions, rather than rare large angle collisions. From \cite{nicholson1983plasma}, the collision frequency $\nu_c$ of a relativistic electron moving at a velocity $v$, due to small angle collisions is:
\begin{equation}\label{collision frequency}
    \nu_c=\frac{8\pi n_e e^4\ln{\Lambda}}{(4\pi\epsilon_0)^2m_e^2v^3\gamma^2}
\end{equation}
Where $\Lambda$ is the plasma parameter and $\gamma$ is the electrons Lorentz factor.\\
Only small-angle collisions are considered, as they primarily propel electrons outward. In contrast, large-angle collisions would significantly alter the electrons' trajectories, leading to a Brownian-like motion rather than directed flow.
\\
Then from \ref{collision frequency}, the mean free path for an electron would be 
\begin{equation}
    \lambda_{mfp}=\frac{(4\pi\epsilon_0)^2m_e^2v^4\gamma^2}{8\pi n_e e^4\ln{\Lambda}}
\end{equation}
Even for electrons with relativistic speeds, the mean free path of an electron would still be much smaller than the average optical depth($\lambda_{mfp}\ll x$).  This implies that the scattered electron will not travel beyond the width $x$, which the photons are reaching. Therefore, the electrons will experience Compton scattering many times in the region within the the reach of the photons and be pushed to the outer edges of the accretion disk. This is why the negatively charged layer's radius can be taken as the width $x$.


\bibliography{sample7}{}
\bibliographystyle{aasjournalv7}



\end{document}